\documentclass[prd,showpacs,twocolumn,aps,superscriptaddress,nofootinbib]{revtex4-1}
\usepackage{amssymb,amsmath,bm}
\usepackage{graphicx}
\usepackage{mathpazo}
\usepackage[colorlinks=true, pdfstartview=FitV, linkcolor=blue, citecolor=blue, urlcolor=blue]{hyperref}

\newcommand{\bx}{{\boldsymbol x}}
\newcommand{\bp}{{\boldsymbol p}}

\newcommand{\bk}{{\boldsymbol k}}

\renewcommand\a{\alpha}
\renewcommand\b{\beta}
\renewcommand\d{\delta}

\renewcommand\l{\lambda}
\renewcommand\r{\rho}

\newcommand\e{\epsilon}
\newcommand\g{\gamma}

\newcommand\m{\mu}
\newcommand\n{\nu}
\newcommand\x{\xi}
\newcommand\p{\pi}

\newcommand\s{\sigma}

\newcommand\ve{\varepsilon}

\renewcommand\S{\Sigma}
\renewcommand\O{\Omega}
\renewcommand\H{\Theta}
\newcommand\D{\Delta}

\newcommand\ra{\rightarrow}

\newcommand\pt{\partial}

\newcommand\lb{\left(}
\newcommand\rb{\right)}
\newcommand\ls{\left[}
\newcommand\rs{\right]}

\newcommand{\lan}{\langle}
\newcommand{\ran}{\rangle}

\newcommand{\non}{\nonumber\\}
\renewcommand{\vec}{\bm}
\newcommand{\mA}{\mathcal{V}}
\newcommand{\tp}{\tilde{p}}
\newcommand{\tD}{\tilde{\D}}
\newcommand{\eq}[1]{Eq.~(\ref{#1})}

\begin{document}

\title{Chiral anomaly in non-relativistic systems: Berry curvature and chiral kinetic theory}
\author{Lan-Lan~Gao}
\affiliation{Physics Department and Center for Particle Physics and Field Theory, Fudan University, Shanghai 200438, China}
\author{Xu-Guang~Huang}
\email{huangxuguang@fudan.edu.cn}
\affiliation{Physics Department and Center for Particle Physics and Field Theory, Fudan University, Shanghai 200438, China}
\affiliation{Key Laboratory of Nuclear Physics and Ion-beam Application (MOE), Fudan University, Shanghai 200433, China}

\begin{abstract}
Chiral anomaly and the novel quantum phenomena it induces have been widely studied for Dirac and Weyl fermions. In most typical cases, the Lorentz covariance is assumed and thus the linear dispersion relations are maintained. However, in realistic materials, such as Dirac and Weyl semimetals, the non-linear dispersion relations appear naturally. We develop a kinetic framework to study the chiral anomaly for Weyl fermions with non-linear dispersions by using the methods of Wigner function and semi-classical equations of motion. In this framework, the chiral anomaly is sourced by Berry monopoles in momentum space and could be enhanced or suppressed due to the windings around the Berry monopoles. Our results can help understand the chiral anomaly-induced transport phenomena in non-relativistic systems.
\end{abstract}
\pacs{11.30.Rd, 72.10.Bg}
\maketitle

\section{Introduction}
During the development of condensed matter physics, the concepts and theories proficient in describing relativistic particles have been often found very useful. Among them, in recent years, the chiral anomaly which was first encountered in particle physics is particularly eye-catching in the studies of three-dimensional topological materials such as Weyl and Dirac semimetals~\citep{Yan:2016euz,Armitage:2017cjs,Lv:2021oam}. In Weyl semimetals, the conduction and valence bands touch at some points (called Weyl nodes) in the Brillouin zone. The Weyl nodes always appear in pairs with opposite chirality around which the low-energy excitations have linear dispersion relations and can be described by Weyl Hamiltonian. In Dirac semimetals, the pairs of Weyl nodes of opposite chirality overlap to form Dirac nodes. The Weyl and Dirac semimetals exhibit a number of novel transport phenomena such as the chiral magnetic effect (CME)~\cite{Kharzeev:2007jp,Fukushima:2008xe} which is a consequence of the chiral anomaly and induces special magnetoresistance~\cite{Li:2014bha,Huang:2015eia,Xiong:2015,Zhang:2016ufu}, and novel modification of Casimir effect~\cite{Fukushima:2019sjn,Rong:2021}. Interestingly, the CME has also been widely discussed in particle and nuclear physics and is under intensive search in relativistic heavy-ion collision experiments~\cite{Kharzeev:2015znc,Huang:2015oca,Hattori:2016emy,Kharzeev:2020jxw}.

Although the Weyl semimetals are considered to be described by Weyl fermions, the realistic excitations in solids can be  different from the ideal Weyl Hamiltonian $\propto\bm\sigma\cdot\bp$ with $\bm\sigma$ the Pauli matrices and $\bp$ the momentum away from the Weyl node. Their dispersion relations typically have tiltings, warpings, and non-linear momentum dependence, which are absent in the ideal Weyl Hamiltonian. These new features may be modeled by an extended Weyl Hamiltonian~\cite{Fang:2012,Soluyanov:2015,Sun:2015,Huang:2017rpa}
\begin{equation}
H=\bm K(\bp)\cdot\bm\s+K_0(\bp) \label{ham0}
\end{equation}
with $K_0$ and $\bm K$ polynomials of $\bp$. A number of interesting phenomena due to the tiltings, warpings, and non-linearity of the Weyl nodes are discussed including, e.g., the modified anomalous Hall effect~\cite{Zyuzin:2016ijd,Nandy:2017knz} and the enhanced CME~\cite{Sharma:2016ero,Yu:2016fvz,Wei:2017tpl}.

Moreover, generalizing the ideal Weyl Hamiltonian to include non-linear momentum dependence is also natural from the view point of low-energy effective theory. The band structure of a real material is usual very complex. Even for the usual Weyl semimetals, the linear Weyl Hamiltonian is considered as a lowest-order effective description valid only near the Weyl nodes. To describe physics away from the Weyl nodes, terms quadratic or even higher-power in momentum need to be included. 

In this paper, we will first focus on a simple example of the Hamiltonian (\ref{ham0}), given by
\begin{equation}
H=\lambda  \vec\sigma\cdot \vec{p} + \alpha \vec{p}^2, \label{ham}
\end{equation}
where $\l=\pm$ denotes the chirality of the Weyl node and the parameter $\a$ depends on the band structure of the material which is assumed to be positive to guarantee the stability of the system. Our purpose is to develop a chiral kinetic theory (CKT) from quantum field theory to describe the semi-classical physics of Hamiltonian (\ref{ham}) and to reveal how the chiral anomaly is self-consistently encoded in such a framework. The underlying methodology is parallel to its relativistic counterpart for relativistic Dirac or Weyl fermions that has been intensively discussed over the past few years~\cite{Gao:2012ix,Chen:2012ca,Huang:2018wdl,Gao:2018wmr,Hidaka:2016yjf,Carignano:2018gqt,Liu:2018xip,Sheng:2020oqs,Huang:2020kik,Hattori:2020gqh,Chen:2021azy}. We then will extend the discussion to the more general Hamiltonian (\ref{ham0}).

The CKT for relativistic systems consists of a set of Lorentz covariant semi-classical kinetic equations that can well describe the anomalous transport effects like the CME and chiral vortical effect (CVE)~\cite{Liu:2018xip,Stephanov:2012ki,Chen:2014cla,Huang:2018aly}. It also provides a convenient tool to explore the topological characteristics of the Dirac or Weyl fermions through the effects of Berry curvature in phase space. For example, the effects of external magnetic or gravitational field and the transports of Dirac fermions in curved space-time can be well expressed through the chiral kinetic theory~\cite{Liu:2018xip,Chen:2014cla,Huang:2018aly}. The relativistic CKT has also been applied to condensed matter systems to study Berry-curvature induced transport phenomena~\cite{Gorbar:2016ygi,Landsteiner:2013sja,Gao:2020gcf}. 

This paper is organized as follows. In section \ref{sec2} , We will derive the CKT for non-relativistic Weyl fermions of type~(\ref{ham}) by using the Wigner function method. We will show that, at the $\hbar$ order (corresponding to one-loop approximation in field-theory calculation), the divergence of the chiral current receives a contribution from the singularity in phase space residing exactly at the Weyl node and gives the chiral anomaly in the same form as derived from relativistic field theory. In section \ref{sec3}, we will directly derive the semi-classical equations of motion for Weyl fermions of type~(\ref{ham0}) and then check the anomaly relation. In section \ref{sec4}, we give comments on our results and discuss potential future application and extension of our results. Throughout this paper, we choose the Fermi velocity $v_F=1$ and the coupling constant $e=1$. To simplify the notations, we will use the four-vector representation like $x^\m=(x_0=t,\bx)$ with index $\m$ from $0$ to $3$, $p^\m=(p_0, \bp)$, $p\cdot x=p_0 t-\bp\cdot\bx$, and use $\eta^{\m\n}=\eta_{\m\n}={\rm diag}(1,-1,-1,-1)$ to raise and lower indices though no Lorentz invariance is assumed. 

\section{Wigner function and chiral anomaly}
\label{sec2}
The Wigner function $W(x,p)$ is a generalization of the classical phase-space distribution function to quantum physics. It is defined as the ensemble average of Wigner operator given by
\begin{equation}
 \label{eq:W}
 \begin{split}
\hat{W}(x,p)=\int d^4y e^{-ip\cdot y/\hbar}{\hat\psi}^\dag(x) e^{y\cdot \overleftarrow{D}^*/2}\otimes e^{-y\cdot D/2}\hat{\psi}(x),
 \end{split}
\end{equation}
where $\hat{\psi}(x)$ is a Weyl spinor operator and $[\hat{\psi}^\dag \otimes \hat{\psi}]_{ab}=\hat{\psi}^\dag_b \hat{\psi}_a$  with $a,b=1$-$2$ being the spinor indices.
Note that we have considered the presence of an external electromagnetic field $A_\mu$ so that $D_\mu$ is the covariant derivative acting on the Weyl spinor:
\begin{equation}
 D_\mu\hat{\psi}(x)
 = \Bigl(
 		\pt_\mu
 		+ iA_\mu/\hbar
   \Bigr) \hat{\psi}(x) \,.
\end{equation}
The Wigner function is given by $W(x,p)=\lan\hat{W}(x,p)\ran$ where $\lan\cdots\ran$ denotes the ensemble average. The equations of motion (Heisenberg equations) of $\hat{\psi}$ and $\hat{\psi}^\dag$ for Hamiltonian (\ref{ham}) are given by:
\begin{eqnarray}
\label{eq:a1}
&iD_t\hat{\psi}+\hbar\alpha
{\bm D}^2\hat{\psi}+i\l{\bm \s}\cdot {\bm D}\hat{\psi}=0,\\
\label{eq:a2}
&i D^*_t\hat{\psi}^\dag-\hbar\alpha{\bm D}^{*2}\hat{\psi}^\dag+i\l {\bm D}^*\hat{\psi}^\dag\cdot{\bm \s}=0.
\end{eqnarray}

We will consider $\hbar$ as being much smaller than the classical action of the system and thus make an expansion in $\hbar$. Combining Eqs.~\eqref{eq:a1}-\eqref{eq:a2} and the definition \eqref{eq:W} for Wigner operator, we derive the equation of motion for Wigner function accurate at $O(\hbar^2)$ as
\begin{equation}
 \begin{split}
  \label{eq:Weq-full}
 &\bigg[\frac{i\hbar}{2}\D_t+p_0+\l\s^i\lb\frac{i\hbar}{2}\D_i+p_i\rb\\&\quad\quad\quad-\a\lb\frac{i\hbar}{2}\D_i+p_i\rb\lb\frac{i\hbar}{2}\D_i+p_i\rb\bigg] W=0,
 \end{split}
\end{equation}
 where $\Delta_\m=\pt_\mu-F_{\mu\n}\pt_p^\n$ (Here, $\D_0=\D_t$) with $F_{\m\n}=\pt_\m A_\n-\pt_\n A_\m$ the field strength, $\pt^\m_p=\pt/\pt p_\m$, and repeated indices are summed. In deriving this equation, we have made an expansion in $\hbar$ with the power counting rule: $p^\m\sim O(1), A_\m\sim O(1)$, and $y^\m\sim i\hbar\pt^\m_p\sim O(\hbar)$, and have used the following equations which are accurate up to $O(\hbar)$:
\begin{eqnarray}
\label{eq:b1}
 &[D_\m,e^{-\frac{y\cdot D}{2}}]=-\frac{iy^\n}{2\hbar} \lb F_{\m\n}-\frac{y^\r}{4}\pt_\r F_{\m\n}\rb e^{-\frac{y\cdot D}{2}}+O(\hbar^2)\,,\non
\label{eq:b4}
&[\pt_\m^y,e^{-\frac{y\cdot D}{2}}]=-\frac{1}{2}e^{-\frac{y\cdot D}{2}}\lb D_\m-\frac{iy^\n}{4\hbar}F_{\m\n}+\frac{iy^\n y^\r}{12\hbar}\pt_\r F_{\m\n}\rb\non&\mkern-120mu+O(\hbar^2)\,.\nonumber
\end{eqnarray}

To reveal the physical content in Eq.~(\ref{eq:Weq-full}), we expand the Wigner function in $su(2)$ algebra, $W=\frac{1}{2}(\mathcal{V}_0+\l\s_i\mathcal{V}_i)$. Note that the Wigner function is Hermitian so that the coefficients $\mathcal{V}_0$ and $\mathcal{V}_i$ are real. The physical meanings of $\mathcal{V}_0$ and $\mathcal{V}_i$ become clear if we express them as $\mathcal{V}_0(x,p)={\rm Tr}\, W(x,p)$ and $\bm{\mathcal{V}}(x,p)=\l{\rm Tr}\,[\bm\s W(x,p)]$. After integration over momentum, $\mathcal{V}_0(x)=\int \frac{d^4p}{(2\p)^4}\mathcal{V}_0(x,p)=\lan\hat{\psi}^\dag(x)\hat{\psi}(x)\ran$~\footnote{In principle, the momentum-space measure should be $d^4p/(2\p\hbar)^4$, but the $\hbar$ here does not influence any of our discussions, so we omit it.} and $\bm{\mathcal{V}}(x)=\l\lan\hat{\psi}^\dag(x)\bm\s\hat{\psi}(x)\ran$. Thus $\mathcal{V}_0(x,p)$ is considered as the phase-space particle number distribution and $\bm{\mathcal{V}}(x,p)$ can be considered as (part of) the particle current (See below) or spin distribution in phase space. These coefficients satisfy the following four equations which can be easily extracted from Eq.~(\ref{eq:Weq-full}):
\begin{eqnarray}
\label{eq:rc1}
&\lb p_0 -\alpha \bp^2+\alpha\frac{\hbar^2}{4}\bm\Delta^2\rb \mathcal{V}^i- p^i\mathcal{V}_0-\l\frac{\hbar}{2}\epsilon^{ijk}
\Delta_j\mathcal{V}^k=0,\;\;\;\;\;\;\;\\
\label{eq:rc2}
&\frac{\hbar}{2}\lb\Delta_t\mathcal{V}^i-2\alpha p_j\Delta_j\mathcal{V}^i+\D_i\mathcal{V}_0\rb+\l\epsilon^{ijk} p_j\mathcal{V}^k=0,\\
\label{eq:rc3}
&\lb p_0 -\alpha \bp^2+\a\frac{\hbar^2}{4}\bm\D^2\rb\mathcal{V}_0+ p_i\mathcal{V}^i=0,\\
\label{eq:kin}
&(\Delta_t-2\alpha p_j \Delta_j)\mathcal{V}_0+\Delta_i\mathcal{V}^i=0.
\end{eqnarray}

We now solve Eqs.~(\ref{eq:rc1})-(\ref{eq:kin}) order by order in $\hbar$. At zeroth order (namely, the classical level), combining  Eqs.~(\ref{eq:rc1})-(\ref{eq:rc3}), one easily finds that
\begin{eqnarray}
\label{zero1}
\mathcal{V}^{(0)}_0&=&4\p \tilde{p}_0 f^{(0)}(x,p)\d(\tilde{p}^2),\\
\label{zero2}
\mathcal{V}^{(0)}_i&=&4\p \tilde{p}_if^{(0)}(x,p)\d(\tilde{p}^2),
 \end{eqnarray}
where we have introduced the shorthand notation $\tilde{p}^\m=(\tilde{p}^0, \tilde{\bp})$ with $\tilde{p}^0=\tilde{p}_0=p_0-\a\bp^2$, $\tilde{p}^i=-\tilde{p}_i=p^i$, and $\tilde{p}^2=\tilde{p}^2_0-\tilde{\bp}^2$. The function $f^{(0)}$ represents the classical distribution function in phase space~\cite{Elze:1986qd,Vasak:1987um}. Substituting $\mathcal{V}^{(0)}_0$ and $\mathcal{V}^{(0)}_i$ into Eq.~(\ref{eq:kin}), we obtain the classical (collisionless) kinetic equation for $f^{(0)}$:
\begin{equation}
\label{classic:kin}
\d(\tilde{p}^2)\ls(p_0-\a\bp^2)(\D_t+2\alpha \bp\cdot\bm\Delta)+\bp\cdot\bm\D\rs f^{(0)}=0.
\end{equation}

To proceed to $O(\hbar)$ order, it is convenient to re-write Eqs.~(\ref{eq:rc1})-(\ref{eq:kin}) at $O(\hbar)$ order in a formally covariant form:
\begin{eqnarray}
\label{eq1:rc1}
&\tilde{p}_\m\mA_\n-\tilde{p}_\n\mA_\m+\l\frac{\hbar}{2}\e_{\m\n\r\s}\tilde{\D}^\r\mA^\s=0,\\
\label{eq1:rc2}
&\tilde{p}\cdot\mathcal{V}=0,\\
\label{eq1:kin}
&\tilde{\Delta}\cdot\mathcal{V}=0,
\end{eqnarray}
where the four-vectors are defined by $\tilde{\D}_\m=(\Delta_t-2\a p_j\D_j, \D_i)$ and $\mathcal{V}^\m=({\mathcal{V}_0,\mathcal{V}^i})$, and the four-dimensional Levi-Civita symbol is normalized to $\e^{0ijk}=-\e_{0ijk}=\e^{ijk}$. The contraction is performed with metric $\eta_{\m\n}$, e.g., $\tilde{p}\cdot\mathcal{V}=\eta_{\m\n}\tilde{p}^\m\mathcal{V}^\n=\tilde{p}_0\mathcal{V}_0-\tilde{\bm p}\cdot\bm\mA$. This set of equations are in the same form as their counterparts in relativistic CKT and thus can be solved in the same way as stressed in Refs.~\cite{Hidaka:2016yjf,Liu:2018xip,Huang:2018wdl}. Multiplying Eq.~(\ref{eq1:rc1}) by an arbitrary vector $n^\m$ satisfying $n^\m n_\m=1$ and $n\cdot \tilde{p}\neq 0$ and combining it with the zeroth order results given in Eqs.~(\ref{zero1})-(\ref{zero2}) and the constraint (\ref{eq1:rc2}), we finally obtain
\begin{eqnarray}
\label{eq:cA}
\mA^\m=4\p\d(\tilde{p}^2)\ls\tilde{p}^\m-\l\frac{\hbar}{\tilde{p}^2}\tilde{G}^{\m\n}\tilde{p}_\n+\l\hbar\S^{\m\n}\tilde{\D}_\n\rs f+O(\hbar^2),\non
\end{eqnarray}
where $\S_{\m\n}=\e_{\m\n\r\s}\tp^\r n^\s/(2\tp\cdot n)$ is called the spin tensor in a frame specified by vector $n^\m$, $f=f^{(0)}+f^{(1)}$ is the distribution function up to order $O(\hbar)$, and $\tilde{G}^{\m\n}=(1/2)\e^{\m\n\r\s} G_{\r\s}$ is the dual tensor of $G_{\m\n}$ which is defined by $G_{\m\n}=-\tilde{\D}_\m\tp_\n$ and related to the field strength tensor via  $G_{0i}=-G_{i0}=F_{0i}-2\a p_jF_{ji}=E^i+2\a\e^{ijk}p^j B^k$ and $G_{ij}=F_{ij}=-\e^{ijk}B^k$ with $\bm E, \bm B$ the electric and magnetic fields. Substituting \eq{eq:cA} into \eq{eq1:kin}, we obtain the $O(\hbar)$-order kinetic equation for $f$:
\begin{equation}
\label{kin:final}
 \begin{split}
&\d(\tilde{p}^2-\l\hbar\S_{\a\b}G^{\a\b})\bigg[ \tp\cdot\tD\\
&\quad\quad+\l\hbar\lb \frac{n_\s}{\tp\cdot n}\tilde{G}^{\s\n}\tD_\n+\tD_\m\S^{\m\n}\tD_\n\rb\bigg] f(t,\bx,p_0,\bp)=0.
 \end{split}
\end{equation}
To derive this equation, we have used the following relations: $\d'(x)=-\d(x)/x$, $\d''(x)=-2\d'(x)/x$, $\tD_\m\tilde{G}^{\m\n}=0$, and $\tilde{G}^{\m\r}G_{\m\s}\tp_\r\tp^\s=\tilde{G}^{\a\b}G_{\a\b}\tp^2/4$ which can be proven using the Schouten identity $\e^{[\m\n\r\s}\tilde{p}^{\l]}=0$ with $[\cdots]$ meaning the anti-symmetrization operation. The form of Eq.~(\ref{kin:final}) shows clearly similarity with its relativistic counterpart in, e.g., Ref.~\cite{Liu:2018xip}.

To see the physical content in Eq.~(\ref{kin:final}) more clearly, it is instructive to choose $n^\m=(1,0,0,0)$ (the laboratory frame). The Dirac $\d$ function in Eq.~(\ref{kin:final}) gives the on-shell condition (dispersion relation) for the fermions in which the term $-\l\hbar\S\cdot G$ expresses a quantum correction due to the coupling between the magnetic moment and magnetic field:
\begin{eqnarray}
\label{eq:dispersion}
p_0=\ve_p=\pm |\bp|(1\mp\hbar\bm \O_\l\cdot\bm B)+\a\bp^2,
\end{eqnarray}
where the upper (lower) sign corresponds to conduction (valance) band of Hamiltonian (\ref{ham}). In the following, we focus on the upper sign. The magnetic moment is determined by a Berry curvature
\begin{eqnarray}
\label{eq:berry}
\bm \O_\l=\l\frac{\bp}{2|\bp|^3}
\end{eqnarray}
of Berry monopole residing at $\bp=\bm 0$ with charge $\l$ (chirality): $\vec\nabla_p\cdot\bm \O_\l=2\p\l\d^{(3)}(\bp)$. Our semi-classical scheme is meaningful only when the quantum correction does not break the band gap $|\bp|$; this requires $\hbar|\bm\O_\l\cdot\bm B|\ll1$ leading to $|\bp|\gg\sqrt{\hbar|\bm B|}$ (the adiabatic condition) at which our kinetic description applies. Integrating Eq.~(\ref{kin:final}) over $\tp_0$ from $0$ to $\infty$, we pickup the contribution of conduction fermions and, after a tedious calculation, find
\begin{equation}
\label{kin:3d}
 \begin{split}
&\Big\{\sqrt{G}\pt_t+\left[\bm v_p+\hbar(\bm v_p\cdot\bm\O_\l)\bm B+\hbar\tilde{\bm E}\times\bm\O_\l\right]\cdot\bm\nabla_x\\
&\quad+\ls\tilde{\bm E}+\bm v_p\times\bm B+\hbar(\tilde{\bm E}\cdot\bm B)\bm \O_\l\rs\cdot\bm\nabla_p\Big\} f(t,\bx,\bp)=0,
 \end{split}
\end{equation}
where $f(t,\bx,\bp)=f(t,\bx,p_0=\ve_p,\bp)$, $\sqrt{G}=1+\hbar\bm\O_\l\cdot\bm B$ is the quantum corrected measure of phase space (see below), $\bm v_p=\bm\nabla_p\ve_p$ is the single-particle velocity, and $\tilde{\bm E}=\bm E-\bm\nabla_x\ve_p$ is a modified electric field. To derive this equation, we have neglected all the $O(\hbar^2)$ terms and used the relations (Note that $\pt_\m$ and $\pt^i_p$ on the left-hand and right-hand sides act on different arguments.): $\pt_\m f(t,\bx,\bp)=\{[\pt_\m +(\pt_\m\ve_p)\pt^0_p]f(t,\bx,p_0,\bp)\}|_{p_0=\ve_p}$ and $\pt^i_p f(t,\bx,\bp)=\{[\pt^i_p +(\pt^i_p\ve_p)\pt^0_p]f(t,\bx,p_0,\bp)\}|_{p_0=\ve_p}$. The kinetic equation for valance fermions can be similarly obtained and the result is in the same form of Eq.~(\ref{kin:3d}) but with $\ve_p$ understood as being given by the lower sign in Eq.~(\ref{eq:dispersion}) and $\l$ replaced by $-\l$ in all the other terms.

By comparing Eq.~(\ref{kin:3d}) with the standard form $(\pt_t+\dot{\bx}\cdot\bm\nabla_x+\dot{\bp}\cdot\bm\nabla_p)f=0$ of Boltzmann equation, we extract the single-particle equations of motion:
\begin{eqnarray}
\label{eq:eom1}
\sqrt{G}\dot{\bx}&=&\bm v_p+\hbar(\bm v_p\cdot\bm\O_\l)\bm B+\hbar\tilde{\bm E}\times\bm\O_\l,\\
\label{eq:eom2}
\sqrt{G}\dot{\bp}&=&\tilde{\bm E}+\bm v_p\times\bm B+\hbar(\tilde{\bm E}\cdot\bm B)\bm \O_\l.
\end{eqnarray}
Similar set of equations of motion has been derived repeatedly in literature in different contexts (though some of them are not complete at $\hbar$ order) for, e.g., band electrons in solids~\cite{Sundaram:1999zz,Xiao:2009rm,Xiao:2005qw,Son:2012wh}, trapped cold atoms~\cite{Huang:2015mga}, neutrinos in supernovae~\cite{Yamamoto:2015gzz}, quarks in quark-gluon plasma~\cite{Stephanov:2012ki,Son:2012zy,Chen:2012ca}, and photons~\cite{Bliokh:2004gz,Onoda:2004zz,Yamamoto:2017uul,Huang:2018aly}. A direct calculation leads to
\begin{eqnarray}
\label{eq:measure}
\pt_t\sqrt{G}+\bm\nabla_x\cdot (\sqrt{G}\dot{\bx})+\bm\nabla_p\cdot(\sqrt{G}\dot{\bp})=2\p\l\hbar\d^{(3)}(\bp)\bm E\cdot\bm B,\non
\end{eqnarray}
where we have used the Maxwell equations $\bm\nabla_x\cdot\bm B=0$ and $\pt_t\bm B=-\bm\nabla_x\times\bm E$ and have omitted $O(\hbar^2)$ terms. This identifies $\sqrt{G}$ as the invariant phase-space measure except for a singularity at $\bp=\bm 0$ at which a Berry monopole resides. This Berry monopole contributes anomalous velocity which could lead to, for example, CME or anomalous Hall effect, and anomalous force, which could lead to chiral anomaly as we show now.

The $U(1)$ Noether current $j^\m=(j^0,j^i)$ associated with Hamiltonian (\ref{ham}) is easily expressed by $\mathcal{V}^\m$,
\begin{eqnarray}
\label{eq:density}
j^0(x)&=&\lan\hat{\psi}^\dag(x)\hat{\psi}(x)\ran=\int \frac{d^4p}{(2\p)^4}\mathcal{V}^0(x,p),\\
\label{eq:current}
j^i(x)&=&\l\lan\hat{\psi}^\dag(x)\s^i\hat{\psi}(x)\ran-i\hbar\a\lan\hat{\psi}^\dag(x)(D_i-\overleftarrow{D}^*_i)\hat{\psi}(x)\ran\non
&=&\int \frac{d^4p}{(2\p)^4}\ls\mathcal{V}^i(x,p)+2\a p^i\mathcal{V}^0(x,p)\rs.
\end{eqnarray}
Substituting Eq.~(\ref{eq:cA}) and after a lengthy but straightforward calculation, we obtain~\footnote{Here, we only present the contributions from conduction fermions. If the contributions of the valance fermions are taken into account, additional terms should be added to $j^0$ and $j^i$ with the same forms as Eqs.~(\ref{eq:density2}) and (\ref{eq:current2}) but with a minus sign for the magnetization current.}
\begin{eqnarray}
\label{eq:density2}
j^0(x)&=&\int \frac{d^3\bp}{(2\p)^3}\sqrt{G}f(t,\bx,\bp),\\
\label{eq:current2}
\bm j(x)&=&\int \frac{d^3\bp}{(2\p)^3}\sqrt{G}\ls\dot{\bx}-\hbar|\bp|\bm\O_\l\times\bm\nabla_x\rs f(t,\bx,\bp).\non
\end{eqnarray}
Note that the second term in Eq.~(\ref{eq:current2}) can be written as $\bm\nabla_x\times\bm M$ with the magnetization $\bm M=\hbar\int \sqrt{G}|\bp|\bm\O_\l f d^3\bp/(2\p)^3$ and thus represents a magnetization current.
Now, using Eqs.~(\ref{kin:3d}) and (\ref{eq:measure}), one can finds that the $U(1)$ current is not conserved at $O(\hbar)$:
\begin{equation}
\label{eq:anomaly}
\pt_\m j^\m=\l\frac{\hbar}{4\p^2}\bm E\cdot\bm B f(t,\bx,\bp=\bm 0).
\end{equation}
If the Fermi surface encloses the point $\bp=\bm0$ so that $f(\bp=\bm0)=1$ at $T=0$, the above relation recover exactly the chiral anomaly relation. Therefore, in the kinetic framework, the chiral anomaly is sourced by the Berry monopole at the Weyl node. Further more, although the presence of $\a\bp^2$ terms in Hamiltonian (\ref{ham}) modifies the single-particle dispersion relation and thus the kinetic equation, it does not change the form of chiral anomaly relation as long as the Fermi surface encloses the Berry monopole.

Before closing this section, we comment on the Wigner-function method in comparison with the methods in ~\cite{Sundaram:1999zz,Xiao:2009rm,Xiao:2005qw,Son:2012wh,Huang:2015mga} based on the wave-packet dynamics of band electrons. Unlike the wave-packet method which can be considered as a bottom-up treatment starting from the single-particle quantum mechanics of the band electrons, the starting point of the Wigner-function method is the underlying quantum field theory. It allows a top-down treatment of the quantum dynamics in a manner of an expansion in $\hbar$. The kinetic equation and the corresponding single-particle equations of motion can be systemcatically obtained to an arbitrary order in $\hbar$ expansion. Moreover, the Wigner-function method can be easily extended to the study of bosons.

\section{Semi-classical equations of motion and Chiral anomaly}
\label{sec3}
The above discusses focus on Hamiltonian (\ref{ham}), we now consider the more general Hamiltonian (\ref{ham0}). We first derive the semi-classical equations of motion of single particle for Hamiltonian (\ref{ham0}). This can be achieved by multiple methods. Here, we use the following well established result~\cite{Stiepan:2013}:\\
Suppose $H(\bx, \bk)=H_0(\bx, \bk)+\hbar H_1(\bx,\bk)+O(\hbar^2)$ with $(\bx,\bk)$ the canonical pairs of mechanical variables. For each isolated eigenvalue $h_0(\bx,\bk)$ of $H_0(\bx,\bk)$, there is associated a classical system with an $\hbar$-dependent Hamilton function $h(\bx,\bk)=h_0(\bx, \bk)+\hbar h_1(\bx, \bk)+O(\hbar^2)$ and a modified symplectic form $G(\bx, \bk) =G_0 +\hbar \Omega(\bx, \bk)+O(\hbar^2)$ ($\O$ is a generalized Berry curvature), where
\begin{eqnarray}
\label{energ}
h_1&=&{\rm tr}\lb H_1\p_0\rb-\frac{i}{2}{\rm tr}\lb\p_0\{\p_0, H_0-h_0\}\rb,\\
\label{berry}
\O_{\a\b}&=&-i{\rm tr}\lb\p_0[\pt_{z^\a}\p_0,\pt_{z^\b}\p_0]\rb.
\end{eqnarray}
Here, the trace is over inner space (for our case, the spin) and $\bx$ and $\bk$ are collectively denoted by $z^\a$ ($\a=1,\cdots, 6$). $\p_0$ is spectral projection to the eigenenergy $h_0$, i.e., $H_0\p_0=h_0\p_0$ and $\p_0^2=\p_0$. The zeroth order symplectic form is given by $G_0=i\s_y\otimes{\bf 1}_3$. The Poisson bracket is defined by $\{A,B\}=\sum_{i=1}^3(\pt_{k^i}A\pt_{x^i}B-\pt_{x^i}A\pt_{k^i}B)$. The commutator is defined by $[A,B]=AB-BA$. If the Hamiltonian depends on time explicitly, $H=H(t,\bx,\bk)$, we just add the canonical pair $(t,E)$ and apply the previous procedure to a new Hamiltonian $\tilde{H}(t,\bx,E,\bk)=E+H(t,\bx,\bk)$. Its spectral projections $\p_0(t, \bx, \bk)$ are independent of $E$ and the classical Hamilton function is
$h(t, \bx,E, \bk) =E +h_0(t, \bx, \bk)+\hbar h_1(t,\bx,\bk)+O(\hbar^2)$ with symplectic form $G = G_0 + \hbar\O(t, \bx, \bp)$, where $h$ and $\O$ are computed from the instantaneous
Hamilton function $H(t, \bx, \bp)$ as before. Note that with symplectic form $G$ the semi-classical equations of motion are
\begin{eqnarray}
\label{semieom}
G_{\a\b}\dot{z}^\b=-\frac{\pt h}{\pt {z^\a}},
\end{eqnarray}
and the invariant phase-space measure is changed to
\begin{eqnarray}
\label{mesureps}
\sqrt{G}\equiv\sqrt{{\rm det}\;(G_{\a\b})}=1+\frac{\hbar}{2}(\O_{x^i k^i}-\O_{k^i x^i})+O(\hbar^2).\non
\end{eqnarray}

For Hamiltonian (\ref{ham0}), $H_1=0$, and $\bp=\bk-\vec A(t,\bx)$ is the kinetic momentum if external electromagnetic field is applied. The eigenenergy and the corresponding projection are
\begin{eqnarray}
\label{energy0}
h_0&=&\pm |\bm K(t,\bx,\bk)|+K_0(t,\bx,\bk)+A_0(t,\bx),\\
\label{proj0}
\p_0 &=&\frac{1}{2}(1\pm \hat{\vec K}\cdot\bm\s),
\end{eqnarray}
where $\hat{\bm K}=\bm K/|\bm K|$. Substituting them into Eqs.~(\ref{energ}) and (\ref{berry}), we find
\begin{eqnarray}
\label{energ1}
h_1&=&\frac{1}{2}\frac{\pt\hat{\bm K}}{\pt k^i}\times\frac{\pt\hat{\bm K}}{\pt x^i}\cdot\bm K,\\
\label{berry1}
\O_{\a\b}&=&\pm\frac{1}{2}\frac{\pt\hat{\bm K}}{\pt z^\a}\times\frac{\pt\hat{\bm K}}{\pt z^\b}\cdot\hat{\bm K},
\end{eqnarray}
and thus the semi-classical equations of motion read
\begin{eqnarray}
\label{xeom2}
\dot{x}^i&=&\lb\d^{ij}+\hbar\O_{k^ix^j}\rb\frac{\pt h}{\pt k^j}-\hbar\O_{k^i k^j}\frac{\pt h}{\pt x^j}+\hbar\O_{k^i t},\\
\label{xeom2}
\dot{k}^i&=&\lb-\d^{ij}+\hbar\O_{x^ik^j}\rb\frac{\pt h}{\pt x^j}-\hbar\O_{x^i x^j}\frac{\pt h}{\pt k^j}-\hbar\O_{x^i t}.
\end{eqnarray}
Using $\bx$ and $\bp$ (the kinetic momentum) as independent variable is more convenient. In this case, the equations of motion (\ref{semieom}) become very simple and take the same form as Eqs.~(\ref{eq:eom1}) and (\ref{eq:eom2}),
\begin{eqnarray}
\label{neweom1}
\sqrt{G}\dot{\bx}&=&\bm v_p+\hbar(\bm v_p\cdot\bm\O)\bm B+\hbar\tilde{\bm E}\times\bm\O,\\
\label{neweom2}
\sqrt{G}\dot{\bp}&=&\tilde{\bm E}+\bm v_p\times\bm B+\hbar(\tilde{\bm E}\cdot\bm B)\bm \O.
\end{eqnarray}
where $\tilde{\bm E}=\bm E +\bm\nabla_x \ve_p$ with $\ve_p=\pm |\bm K(\bp)|+K_0(\bp)+h_1(\bx,\bp)$, ${\bm v_p}=\bm\nabla_p\ve_p$, and $\O^i=\frac{1}{2}\e^{ijk}\O_{p^jp^k}$. With these equations of motion, the collisionless kinetic equation can be written as $(\pt_t+\dot{\bx}\cdot\bm\nabla_x+\dot{\bp}\cdot\bm\nabla_p)f=0$ whose explicit form is same as Eq.~(\ref{kin:3d}):
\begin{equation}
\label{kin:3d2}
 \begin{split}
&\Big\{\sqrt{G}\pt_t+\left[\bm v_p+\hbar(\bm v_p\cdot\bm\O)\bm B+\hbar\tilde{\bm E}\times\bm\O\right]\cdot\bm\nabla_x\\
&\quad+\ls\tilde{\bm E}+\bm v_p\times\bm B+\hbar(\tilde{\bm E}\cdot\bm B)\bm \O\rs\cdot\bm\nabla_p\Big\} f(t,\bx,\bp)=0.
 \end{split}
\end{equation}

Let us seek for an action in the form $S=\int dt[U^i(t,\bx,\bp)\dot{x}^i+W^i(t,\bx,\bp)\dot{p}^i-\x(t,\bx,\bp)]$ for Eqs.~(\ref{neweom1}) and (\ref{neweom2}). Using the least action principle, we determine $U^i, W^i$, and $\x$ as $U^i=p^i+A^i(t,\bx)$, $W^i=-\hbar a^i(\bp)$, and $\x=\ve_p+A_0(t,\bx)$:
\begin{eqnarray}
\label{action}
S=\int dt\ls (\bp+\bm A(t,\bx))\dot{\bx}-\hbar\bm a(\bp)\cdot{\bp}-\ve_p-A_0(t,\bx)\rs,\non
\end{eqnarray}
where $\bm a(\bp)$ is determined by the condition $\bm\nabla_p\times\bm a=\bm\O$ so that $\bm a$ is the Berry connection. Using $S$ we obtain the number density and current as
\begin{eqnarray}
\label{covcurrent}
j^\m(x)&=&-\int \frac{d^3\bm y d^3\bp}{(2\p)^3}\sqrt{G}\frac{\d S(\bm y,\bp)}{\d A_\m(t,\bx)}f(t,\bx,\bp).
\end{eqnarray}
The results are
\begin{eqnarray}
\label{Sdensity} 
j^0(x)
&=&\int \frac{d^3\bp}{(2\p)^3}\sqrt{G}f(t,\bx,\bp),\\
\label{Scurrent} 
\bm j(x)
&=& \int \frac{d^3\bp}{(2\p)^3}\sqrt{G}\ls\dot{\bx}\mp\hbar|\bm K(\bp)|\bm\O\times\bm\nabla_x\rs f(t,\bx,\bp),\non
\end{eqnarray}
where the second term in $\bm j$ is because $h_1(\bx,\bp)$ depends on $\bm A$. Note the similarity with equations (\ref{eq:density2}) and (\ref{eq:current2}). The divergence of $j^\m$ reads (omitting $O(\hbar^2)$ terms)
\begin{eqnarray}
\label{anom_current}
\pt_\m j^\m&=&\int \frac{d^3\bp}{(2\p)^3}\ls\pt_t\sqrt{G}+\bm\nabla_x\cdot (\sqrt{G}\dot{\bx})+\bm\nabla_p\cdot(\sqrt{G}\dot{\bp})\rs f\non
&=&\hbar\bm E\cdot\bm B\int \frac{d^3\bp}{(2\p)^3}\bm\nabla_p\cdot\bm \O(\bp) f(t,\bx,\bp),
\end{eqnarray}
where we have used~\footnote{In terms of the generalized Berry curvature $\O_{\a\b}$, it can be expressed by $\frac{\pt\sqrt{G}}{\pt t}+\frac{\pt\sqrt{G}\dot{x}^i}{\pt x^i}+\frac{\pt\sqrt{G}\dot{p}^i}{\pt p^i}=\Theta_{p^i x^i t}+\Theta_{p^jx^jx^i}\pt_{p^i}\varepsilon_p
+\Theta_{x^j p^jp^i}\pt_{x^i}\ve_p$ where $\H_{\a\b\g}=\pt_\a\O_{\b\g}+\pt_\b\O_{\g\a}+\pt_\g\O_{\b\a}$ is the Berry monopole charge function which is the exteriror derivative of Berry curvature~\cite{Gao:2020gcf}.}
\begin{eqnarray}
\label{eqmeasure}
\pt_t\sqrt{G}+\bm\nabla_x\cdot (\sqrt{G}\dot{\bx})+\bm\nabla_p\cdot(\sqrt{G}\dot{\bp})&=&\hbar\bm\nabla_p\cdot\bm \O(\bp)\tilde{\bm{E}}\cdot\bm B, \non
\end{eqnarray}
which is a direct consequence of Eqs.~(\ref{neweom1}) and (\ref{neweom2}). The quantity $\vec\nabla_p\cdot\vec\O$ vanishes except for points where $\vec\O$ is singular. In fact, substituting the expression for $\vec\O$, we have
\begin{eqnarray}
\label{monopole}
\bm\nabla_p\cdot\bm \O(\bp)
&=&{\rm det}\lb\frac{\pt K^a}{\pt p^i}\rb\vec\nabla_K\cdot\vec\O_K\non
&=&\pm 2\pi\,{\rm det}\lb\frac{\pt K^a}{\pt p^i}\rb\d^{(3)}(\vec K),
\end{eqnarray}
where the repeated indices all summed over $1-3$ and $\vec\O_K=\pm\hat{\vec K}/(2\bm K^2)$ is the Berry curvature in $\vec K$-space. To derive this relation, we have used the Schouten identity $\e^{ijk}K^{[a}_{,j}K^b_{,k}K^d_{,i}\O_K^{c]}=0$ with $K^a_{,j}=\pt K^a/\pt p^j$. The determinant in the right-hand side of Eq.~(\ref{monopole}) represents the Berry monopole charge in $\vec K$-space and is mathematically a Jacobian for the map from $\vec p$-space to $\vec K$-space which makes the integral of $\vec\nabla_p\cdot\vec\O$ have clear topological meaning:
\begin{eqnarray}
\label{monopoleintegral}
\frac{1}{2\p}\int d^3\bp \bm\nabla_p\cdot\bm \O(\bp)&=&\pm \sum_{\vec p^*}N_{\bp^*},
\end{eqnarray}
where $\bp^*$ is the points in $\bp$-space where $\vec K(\bp^*)=0$ (i.e., the location of Berry monopole in $\bp$-space) and $N_{\bp^*}$ is the winding number of map $\bp\rightarrow\vec K=\vec K(\bp)$ around $\bp^*$. Collecting the above results, we obtain the following relation expressing the chiral anomaly:
\begin{eqnarray}
\label{anom_current2}
\pt_\m j^\m&=&\pm\frac{\hbar}{4\p^2}\bm E\cdot\bm B\sum_{\vec p^*}N_{\bp^*} f(t,\bx,\bp^*).
\end{eqnarray}
This extends Eq.~(\ref{eq:anomaly}) to a more general case described by Hamiltonian (\ref{ham0}). It shows that the term $K_0$ in Hamiltonian (\ref{ham0}) does not change the form of the anomaly relation and contributes to the chiral anomaly only through the distribution function $f$. We emphasize that the relation (\ref{anom_current2}) can also be understood through the index theorem~\cite{Yee:2019rot}.

To end this section, we give the expressions for CME of fermions of type (\ref{ham0}) for an equilibrium distribution function which depends only on the energy $\ve_p$, $f=f(\ve_p)$,
\begin{eqnarray}
\label{cmecurrent}
\bm j_{\rm CME}&=&\mp\frac{\hbar}{4\p^2}\bm B\sum_{\vec p^*}N_{\bp^*} F(\ve_{\bp^*}),
\end{eqnarray}
where $F({\ve_p})$ satiesfies $F'(\ve_p)=f(\ve_p)$ and the boundary condition $F(\ve_p)\ra0$ for $\bp\ra\infty$.

\section{Summary and outlook}
\label{sec4}
In summary, we have thoroughly examined chiral anomaly for fermions described by Hamiltonians (\ref{ham0}) and (\ref{ham}) via the kinetic approach. For Hamiltonian (\ref{ham}), we use the  Wigner function method and derive the corresponding chiral kinetic equation up to $O(\hbar)$ order. Comparing to the relativistic case, the only change is the on-shell condition which constrains also the single-particle equations of motion and the distribution function. The chiral anomaly is seen to be sourced by the Berry monopole at $\bp=\bm 0$ and is formally unchanged comparing to the relativistic case. We then extend the analysis to the more general Hamiltonian (\ref{ham0}) by directly applying the semi-classical expansion to it. This amounts to map the $O(\hbar)$ order behavior of the quantum dynamics of a spinful particle to the classical dynamics of a spinless particle equipped with a modified $\hbar$-dependent Hamilton function and sympletic form. The Berry monopole charge is given by the winding number of the map from $\bp$-space to $\bm K$-space around each Berry monopole. Due to this, the chiral anomaly relation is correspondly changed [see Eq. (\ref{anom_current2})].

From the above analysis, the source of chiral anomaly from the point of view of kinetic theory is clearly displayed, showing that the chiral anomaly has an infrared origin stemming from the singularities in phase space (the Berry monopole). This could be helpful for understanding the related topological transport phenomena in non-relativistic systems. It would be also interesting to make a field-theoretical analysis for Hamiltonians (\ref{ham0}) and (\ref{ham}) to reveal the connection of chiral anomaly with ultraviolet divergence in field theory. Such an analysis has been performed for Hamiltonian (\ref{ham0}) without $K_0$ term~\cite{Yee:2019rot} and will be presented for the case with $K_0$ in a future work.

\begin{acknowledgments}
The authors thank Sahal Kaushik, Dmitri Kharzeev, Yu-Chen Liu, Satoshi Nawata, Evan Philip, Xin-Li Sheng, and Yong-Shi Wu for helpful discussions. This work is supported by NSFC under Grant No.~12075061 and Shanghai NSF under Grant No.~20ZR1404100.
\end{acknowledgments}

\bibliography{bibfile}

\end{document}